\documentclass[adp,fleqn]{w-art}
\usepackage{times}
\usepackage{w-thm}
\usepackage{amsmath,amssymb}
\usepackage[]{graphicx}
\chardef\bslash=`\\ 

\newcommand{\Na}[1]{{\Hat{#1}}}
\newcommand{\Fe}[1]{{\bf #1}_{F}}
\newcommand{\Fep}[1]{{\bf #1}'_{F}}
\newcommand{\Avp}[1]{\left\langle{#1}\right\rangle_{\Fe p}}
\newcommand{\Avpp}[1]{\left\langle{#1}\right\rangle_{\Fep p}}

\begin{document}
\DOIsuffix{theDOIsuffix}
\Volume{12}
\Issue{1}
\Copyrightissue{01}
\Month{01}
\Year{2003}
\pagespan{1}{}



\title[$d$-wave superconductors near boundaries]
{Influence of surface roughness on subdominant pairing\\ in $d$-wave superconductors}


\author[T.\ L\"uck]{Thomas L\"uck\footnote{E-mail: {\sf lueckt@physik.uni-augsburg.de}, Phone: +49\,821\,598\,3255,
     Fax: +49\,821\,598\,3262}\inst{1}} 
\address[\inst{1}]{Universit\"at Augsburg, Institut f\"ur Physik, Theoretische Physik II, 86135 Augsburg, Germany}

\begin{abstract}
  We study a $d$-wave superconductor with dominant $d_{x^2-y^2}$-wave
  order parameter and subdominant pairing in either the $s$- or the
  $d_{xy}$-wave channel near a surface. In particular we analyze the
  influence of surface roughness on the mixed order parameter which
  may break the time-reversal symmetry. We find that the subdominant
  component is suppressed by the roughness independent of its pairing
  symmetry; for very rough surfaces the subdominant component may even
  vanish completely.  Additionally we discuss a possible real-valued
  admixture which counteracts the suppression of the
  $d_{x^2-y^2}$-wave order parameter at the surface.
\end{abstract}
\maketitle                   





There is an on-going discussion about the surface state of high-$T_c$
materials in the superconducting phase, which is commonly believed to
exhibit $d$-wave symmetry in the bulk. In particular the question is
whether a subdominant order parameter occurs at the surface which
leads to a breaking of the time-reversal symmetry. Theoretical studies
show that this is possible for low enough temperatures if the $d$-wave
order parameter is strongly suppressed at the surface, and a
subdominant pairing interaction exists~\cite{Ma95,Ma951,Fo97}: Such a
situation is realized, for example, if the surface is oriented in the
$[110]$ direction of a $d_{x^2-y^2}$-wave superconductor with
subdominant pairing channels of the $s$-wave or the $d_{xy}$-wave
type. This would manifest itself in the differential conductance of
tunnel junctions as a splitting of the zero-bias peak which occurs for
a tilted single-component $d_{x^2-y^2}$-wave order
parameter~\cite{Fo97}. Another experimental possibility is to study
the magnetic field dependence of the differential conductance, which
would also show anomalies due to a subdominant pairing
interaction~\cite{Fo03}. In this work we concentrate on the zero-field
case.

In some experiments on high-$T_c$ superconductors a splitting of the
zero-bias conductance peak was observed~\cite{Co97,Da01}, which seems
to confirm the existence of a subdominant order parameter. On the
other hand, in numerous similar experiments no such splitting was
visible~\cite{Al97,Ig00,Ye01}.  It was shown experimentally for
YBCO~\cite{Da01} that one reason for this discrepancy might be a
different oxygen doping, i.e.  different electronic properties of the
material.  We show here that these results can also be explained by
surface roughness, as a rough surface can suppress a subdominant
admixture of either $d_{xy}$-wave or $s$-wave symmetry. Moreover we
demonstrate that in some cases a subdominant admixture to the order
parameter occurs without breaking the time-reversal symmetry; then the
spatial symmetry is modified near the surface.

To study the superconducting phase of high-$T_c$ superconductors we
start with a few symmetry considerations.  In conventional
superconductors the gauge symmetry is broken only, and usually the
order parameter is assumed to be isotropic ($\Delta(\Fe p)=\Delta$,
$\Fe p$: Fermi momentum).  In unconventional superconductors the
lattice symmetry is broken in addition to the gauge symmetry, and the
order parameter is anisotropic. In order to describe high-$T_c$
materials we consider a tetragonal lattice symmetry; note that we
neglect possible orthorhombic distortions, present for example in
YBCO. Then the order parameter can be expanded as
\begin{equation}
\Delta(\Fe p)=-\sum_i\Delta_i\eta_i(\Fe p)\;,
\end{equation}
where $\eta_i$ are basis functions of the irreducible representations
of the tetragonal space group, $D_{4h}$. The lowest-order basis
functions are given in Table~\ref{reps}. Since we neglect
the weak dispersion in $c$-direction, we concentrate on three order
parameter symmetries, namely $d_{x^2-y^2}$-, $d_{xy}$- and
$s$-wave (see Fig.~\ref{basis}).
\begin{vchtable}[b]\noindent
\begin{tabular}{p{1.2cm} p{3.6cm} p{3cm} }
$D_{4h}$: &basis functions & \\
\hline
$A_{1g}$ & $\eta_1({\bf p})=1,p_x^2+p_y^2,p_z^2$ & (anisotropic) $s$-wave\\
$A_{2g}$ & $\eta_2({\bf p})=p_xp_y(p_x^2-p_y^2)$ &\\
$B_{1g}$ & $\eta_3({\bf p})=p_x^2-p_y^2$ & $d_{x^2-y^2}$-wave\\
$B_{2g}$ & $\eta_4({\bf p})=p_xp_y$ & $d_{xy}$-wave\\
$E_g$ & $\eta_5({\bf p})=(p_xp_z,p_yp_z)$ \\
\end{tabular}
\caption{\label{reps}Basis functions of lowest order for the group
$D_{4h}$.}
\end{vchtable}
\begin{vchfigure}[b]
  \includegraphics[width=0.75\textwidth]{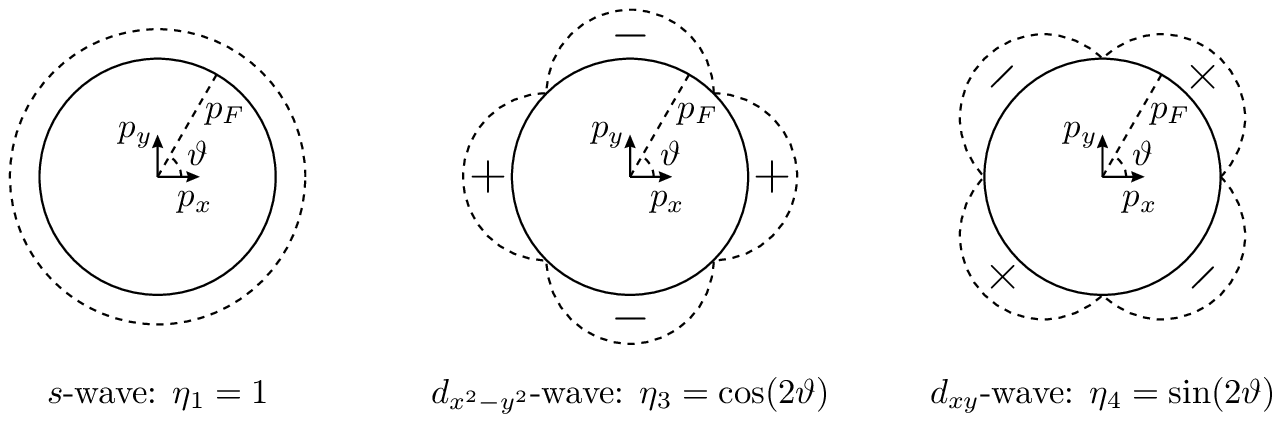}
  \vchcaption{\label{basis}Angular dependence of the order parameter
    for three different symmetries: $s$-, $d_{x^2-y^2}$-, and
    $d_{xy}$-wave.}
\end{vchfigure}
In the following we will discuss a dominant $d_{x^2-y^2}$-wave order
parameter, and a weaker pairing interaction in either the $s$-wave or
the $d_{xy}$-wave channel.

Therefore we consider a superconductor with two (or more) possible
order parameter components. Due to the pairing interaction each
component has its individual critical temperature, $T_{c,i}$; i.e.  in
a bulk superconductor the $i^{\rm th}$ order parameter component would
become finite at $T_i$ if all other components were zero.  The two
cases of interest for us, namely a $d_{x^2-y^2}$-wave order parameter
with $s$- or $d_{xy}$-wave admixture, are discussed in
Refs.~\cite{Be98} and~\cite{Ba97}, respectively, with the following
results: If both components, the dominant and the subdominant, have a
finite critical temperature ($T_{c,i}>0$), it is possible that the
dominant order parameter component with $T_{c,1}$ suppresses the
subdominant component, which then becomes finite at a lower
temperature $\tilde T_{c,2}<T_{c,2}<T_{c,1}$; alternatively it may
happen that the subdominant pairing channel is completely suppressed,
i.e. $\tilde T_{c,2}=0$. The latter situation can be assumed for
high-$T_c$ superconductors, where in the bulk only a
$d_{x^2-y^2}$-wave order parameter is observed, while other order
parameter components seem to be absent.

Although the subdominant pairing can be neglected in the bulk, it
might become important near inhomogeneities, e.g. impurities or
boundaries, which tend to suppress the dominant $d_{x^2-y^2}$-wave order
parameter. At a specular surface, for example, a $d_{x^2-y^2}$-wave
order parameter which is tilted by $45^\circ$ with respect to the
surface normal is suppressed completely. In the following we will
discuss the interplay between the suppression of the
$d_{x^2-y^2}$-wave order parameter near an interface and the existence
of a subdominant order parameter component in this region. In
particular we study the influence of interface roughness, which also
modifies the behavior of the $d_{x^2-y^2}$-wave component.

Here we apply the theory of quasiclassical Green's functions in
thermal equilibrium~\cite{Ei68,RaSm86}. The Green's functions are
determined by the Eilenberger equation,
\begin{equation}\label{Eilen_eq}
\left[\Na\tau_3E+e\Na\tau_3\Fe v\cdot{\bf A}({\bf r})+
i\Na\Delta(\Fe p,{\bf r}),\Na g(E,\Fe p;{\bf r})\right]+
i\Fe v\cdot\partial_{\bf r}\Na g(E,\Fe p;{\bf r})=0\;,
\end{equation}
supplemented by the normalization condition
\begin{equation}\label{norm}
\left[\Na g(E,\Fe p;{\bf r})\right]^2=\Na 1\;.
\end{equation}
Here $\hat\tau_i$ are the Pauli-matrices in Nambu-space. The
(spin-singlet) order parameter
\begin{equation}
\hat \Delta(\Fe p,{\bf r})=
\begin{pmatrix}
0 & \Delta(\Fe p,{\bf r})\\ \Delta^*(\Fe p,{\bf r})& 0
\end{pmatrix}\;
\end{equation}
must be determined self-consistently via
\begin{equation}\label{OP}
\Na\Delta(\Fe p,{\bf r})=-\pi{\cal N}_0  T\sum_{|E_n|<E_c}
\Avpp{V(\Fe p,\Fep p)\Na g(iE_n,\Fep p;{\bf r})}\;,
\end{equation}
where $\Avpp{\dots}$ denotes an average over the Fermi surface
(assumed to be spherical).  The cut-off energy is $E_c$, and the
normal-state density of states at the Fermi energy is denoted by
${\cal N}_0$ per spin.  According to the above remarks on the order
parameter symmetry we can expand the pairing interaction as follows:
\begin{equation}
V(\Fe p,\Fep p)=-\sum_iV_i\eta_i(\Fe p)\eta_i(\Fep p)\;.
\end{equation}
The current density can be determined from the Greeen's function via
\begin{equation}
{\bf j}({\bf r})=-ie\pi{\cal N}_0T\sum_{E_n}
{\rm Tr}\left[\Na\tau_3\Avp{\Fe v\Na g(iE_n,\Fe p;{\bf r})}\right],
\end{equation}
and the angle-resolved density of states is given by
\begin{equation}
{\cal N}(E,\Fe p;{\bf r})=\frac{{\cal N}_0}{2}{\rm Re}
\Big\{{\rm Tr}\left[\Na\tau_3\Na g(E+i0_+,\Fe p;{\bf r}) \right]\Big\}.
\end{equation}
The density of states at the surface is connected to the differential
tunnel conductance by the relation~\cite{TL01}
\begin{equation}
G(V)=Ae^2\Avp{v_{F,x}{\cal T}(\Fe p){\cal N}(eV,\Fe p;0)},
\end{equation}
where ${\cal T}(\Fe p)={\cal T}_0p^2_{F,x}/p_F^2$ (${\cal T}_0\ll 1$)
is the angle-dependent transparency, and $A$ is the cross-section of
the tunneling contact; this results in the normal state resistance
$R_N^{-1}=4Ae^2{\cal N}_0v_F{\cal T}_0/3\pi$.

\begin{vchfigure}[htb]
  \includegraphics[width=0.7\textwidth]{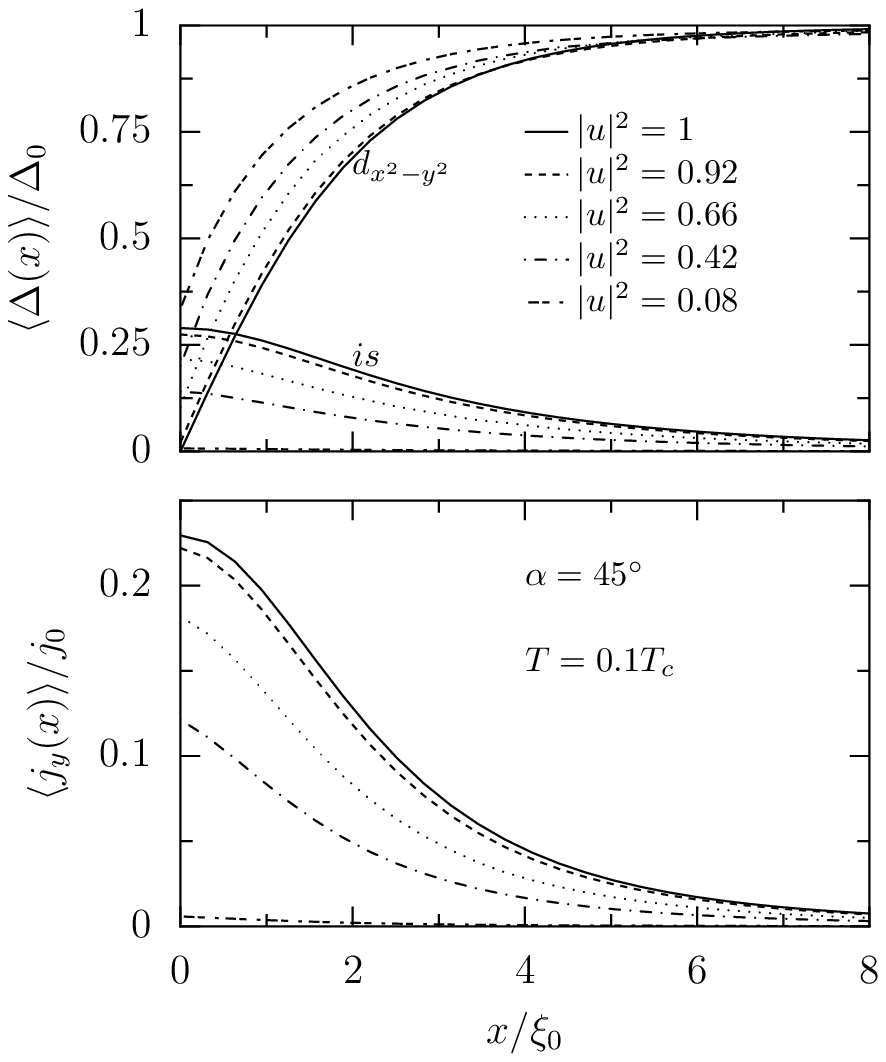}
  \vchcaption{\label{opcuds45} Dominant $d_{x^2-y^2}$-wave order
    parameter with (imaginary) subdominant $s$-wave component, for
    $\alpha=45^\circ$ at $T=0.1T_c$ (upper panel), and the resulting
    current flowing parallel to the surface (lower panel);
    $\Delta_0=2.14T_c$, $\xi_0=v_F/\pi\Delta_0$, and $j_0=2e{\cal
      N}_0v_F\Delta_0$.}
\end{vchfigure}

The quasiclassical approach is valid only on scales which are large
compared to the Fermi wavelength, and is not directly applicable near
interfaces.  This problem must be circumvented using boundary
conditions for the Green's functions. For specular interfaces, i.e.
when the momentum parallel to the interface is conserved, one can use
Zaitsev's boundary conditions~\cite{Za84}; these can be extended to
include interface roughness~\cite{Ya96,BaSv97,Po99,Po00}. We use here
an approach which was introduced by Shelankov and Ozana~\cite{ShOz00},
in which the properties of the interface are described by a scattering
matrix.

\begin{vchfigure}[t]
  \includegraphics[width=0.7\textwidth]{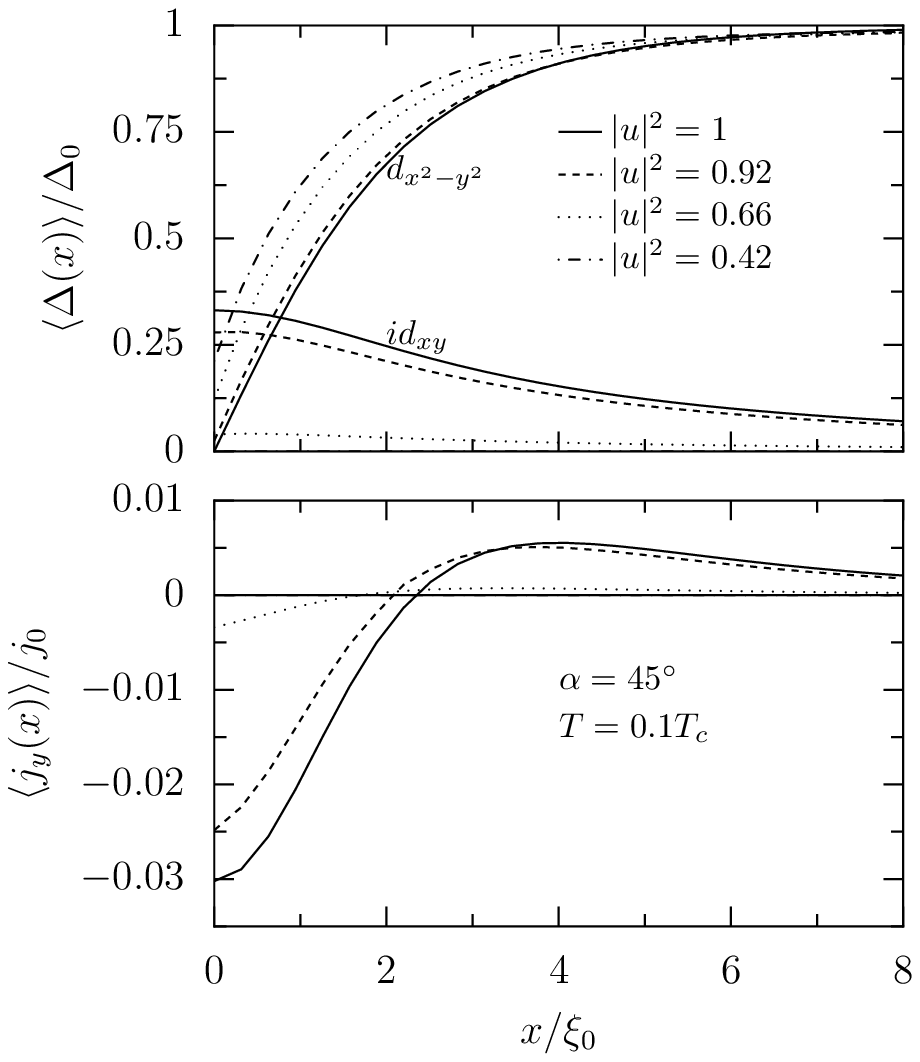}
  \vchcaption{\label{opcudd45}Dominant $d_{x^2-y^2}$-wave order
    parameter with (imaginary) subdominant $d_{xy}$-wave component,
    for $\alpha=45^\circ$ at $T=0.1T_c$ (upper panel), and the
    resulting current flowing parallel to the surface (lower panel);
    $\Delta_0=2.14T_c$, $\xi_0=v_F/\pi\Delta_0$, and $j_0=2e{\cal
      N}_0v_F\Delta_0$.}
\end{vchfigure}

In order to describe interface roughness, which is present on the
atomic scale, we choose the scattering matrix to be a random unitary
matrix~\cite{Ya96,TL01,TL03}. Here the only relevant roughness
parameter is the probability for specular scattering, $|u|^2$.  A
rough interface leads to a reduced weight for specular scattering,
$|u|^2<1$, and with a probability of $1-|u|^2$ a particle is scattered
randomly into any other direction. This model is described in detail
in~\cite{TL01,TL03}. The disorder averaged quantities are denoted by
$\langle\dots\rangle$.

The superconductor is described by a dominant $d_{x^2-y^2}$-wave order
parameter with a critical temperature $T_c$, and a second order
parameter component of either $s$- or $d_{xy}$-wave type with an
individual critical temperature $T_{c,2}=0.3T_c$; this choice ensures
that in the bulk only a $d_{x^2-y^2}$-wave order parameter is present.
The surface normal is in the $x$-direction, and we assume translational
invariance in the $y$-direction (parallel to the surface).

The behavior of a single-component $d_{x^2-y^2}$-wave superconductor
near a surface is well known~\cite{Ya96,BaSv97,Po99,Po00,TL01}. The
following results are of particular importance for our purpose: If the
order parameter is tilted with respect to the surface normal,
trajectories exist where the order parameter changes its sign due to
surface scattering; this leads to a zero-bias peak in the differential
conductance of tunnel contacts, and a suppression of the order
parameter near the surface. Without surface roughness and for a
tilting angle of $45^\circ$ the order parameter at the surface is
zero. Surface disorder leads to a finite value of the order parameter,
and in most models to a broadened zero-bias conductance peak.
Furthermore, even an untilted $d_{x^2-y^2}$-wave order parameter is
suppressed near a rough surface~\cite{TL01}.

\begin{vchfigure}[htb]
  \includegraphics[width=0.7\textwidth]{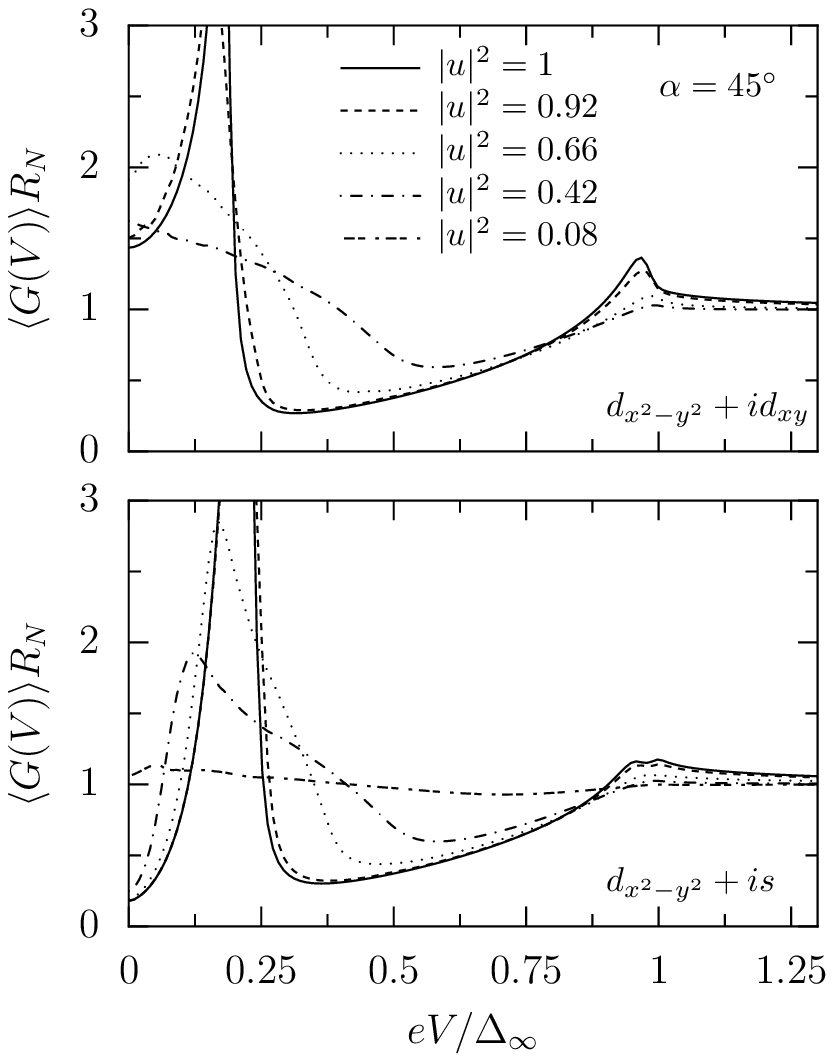}
  \vchcaption{\label{dos45}Differential conductance versus voltage for
    $\alpha=45^\circ$, for subdominant pairing in the $s$-wave (lower
    panel) and $d_{xy}$-wave channel (upper panel).}
\end{vchfigure}

First we focus on a $d_{x^2-y^2}$-wave superconductor with a
subdominant pairing in the $s$-wave channel; the order parameter is
rotated by $45^\circ$ with respect to the surface normal. Here it is
important to note that an isotropic $s$-wave order parameter itself is
inert against scattering at a rough surface. Without roughness
($|u|^2=1$) the dominating $d_{x^2-y^2}$-wave component is completely
suppressed at the surface ($x=0$), and a finite $s$-wave admixture
occurs near the boundary with a relative phase $\varphi=\pm\pi/2$
(Fig.~\ref{opcuds45}). This state has a spontaneously broken
time-reversal symmetry near the surface.  As a consequence a current
flowing in $y$-direction is present (Fig.~\ref{opcuds45}), and a
splitting of the zero-bias conductance peak is found
(Fig.~\ref{dos45}). These are well-known results for the clean
case~\cite{Ma95,Ma951,Fo97,Ra98}.

For a rough surface the $d_{x^2-y^2}$-wave component recovers and has
a finite value at the surface; this in turn leads to a suppression of
the subdominant component, the suppression becoming stronger with increasing
roughness (Fig.~\ref{opcuds45}). As a consequence the splitting of the
zero-bias conductance peak and the parallel current are reduced. For
very strong roughness ($|u|^2\lesssim 0.08$) no subdominant order
parameter is observed. Therefore the differential conductivity is
no longer affected by the subdominant pairing, and no parallel current
exists.

\begin{vchfigure}[htb]
  \includegraphics[width=0.7\textwidth]{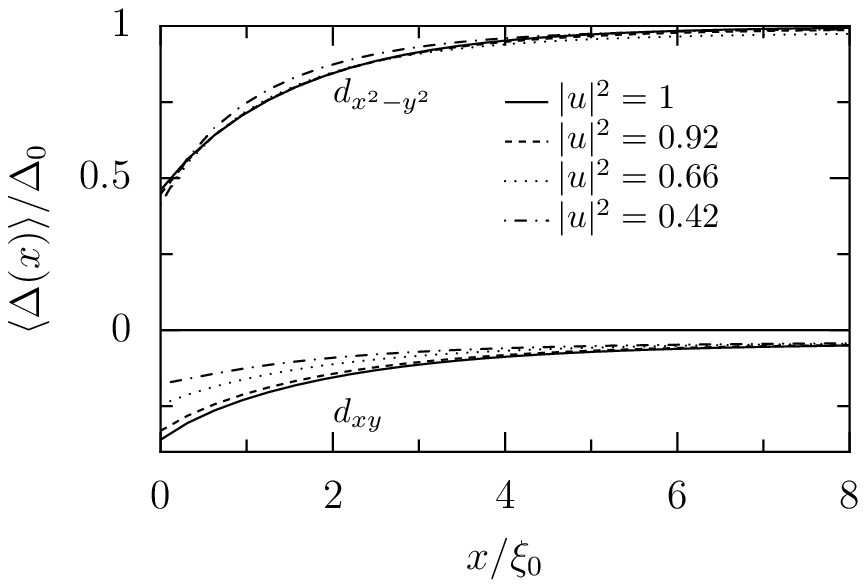}
  \vchcaption{\label{opcudd24}Dominant $d_{x^2-y^2}$-wave order
    parameter with (real) subdominant $d_{xy}$-wave component, for
    $\alpha=24^\circ$ at $T=0.1T_c$; $\Delta_0=2.14T_c$, and
    $\xi_0=v_F/\pi\Delta_0$.}
\end{vchfigure}
\begin{vchfigure}[htb]
  \includegraphics[width=0.7\textwidth]{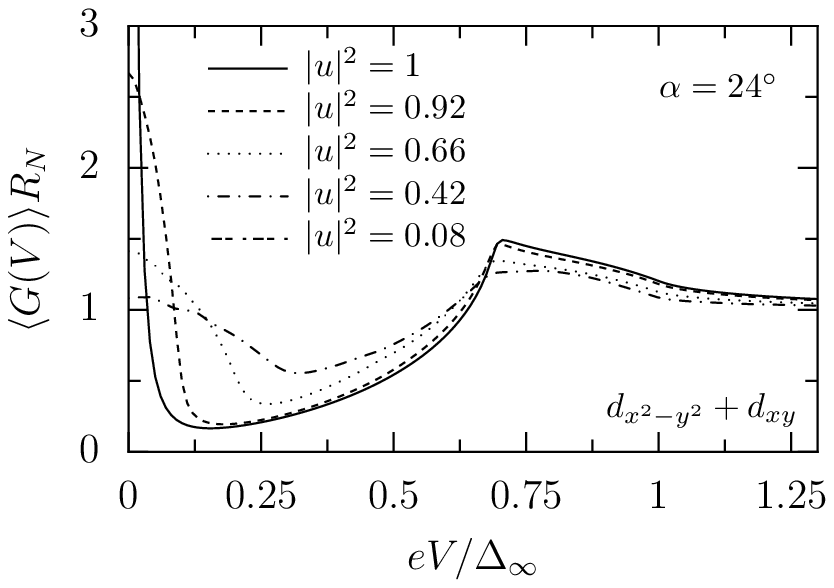}
  \vchcaption{\label{dosdd24}Differential conductance versus voltage
    for $\alpha=24^\circ$, for subdominant pairing in the
    $d_{xy}$-wave channel. Note, that for the clean case ($|u|^2=1$)
    the zero-bias conductance peak has been broadend by adding a
    finite imaginary part to the energy.}
\end{vchfigure}

Next we consider the situation with a subdominant
$d_{xy}$-wave component. In the clean case we find similar results
as above, since the $45^\circ$-rotated $d_{xy}$-wave order parameter
is not suppressed by the surface. Therefore a finite $d_{xy}$-wave
admixture with a relative phase of $\varphi=\pm\pi/2$ exists near the
surface, and the resulting state exhibits broken time-reversal
symmetry. Consequently a current in $y$-direction flows near the
surface~(Fig.~\ref{opcudd45}), and the zero-bias conductance peak
splits (Fig.~\ref{dos45}).

Here the current behaves qualitatively different (Fig.~\ref{opcuds45}) to what
we found for subdominant $s$-wave pairing (Fig.~\ref{opcudd45}); in particular
here the current changes its sign at some distance from the surface.
This is related to the fact that a mixed $d_{x^2-y^2}+id_{xy}$ order
parameter has a finite magnetic moment, which is not the case for a
$d_{x^2-y^2}+is$ order parameter.

A crucial difference occurs for rough surfaces: The $d_{xy}$-wave
component is not only suppressed by the dominant order parameter but,
as it is anisotropic, also by the surface roughness.  Therefore the
subdominant admixture decreases rapidly with increasing surface
roughness; we find a full suppression already for $|u|^2\lesssim 0.42$
(Fig.~\ref{opcudd45}).

We now study the situation when the order parameter is rotated by only
$24^\circ$ with respect to the surface normal, and a subdominant
$d_{xy}$-wave pairing is present. The crucial difference to the
previous cases is that the dominant order parameter is not completely
suppressed by the specular surface. We also find a finite value of the
subdominant order parameter but with a relative phase $\varphi\approx
0$; i.e. the admixture is mainly real-valued, and the time-reversal
symmetry-breaking imaginary part is negligible (Fig.~\ref{opcudd24}).
Therefore no splitting of the zero-bias conductance peak can be
observed (Fig.~\ref{dosdd24}). The real-valued subdominant component
leads to an effective rotation of the dominant order parameter such
that the tilting angle is diminished.  As a result the suppression
of the order parameter at the surface is reduced; the
differential conductance shows that spectral weight is shifted from
lower to higher energies compared to the case without subdominant
admixture.  For subdominant $s$-wave pairing a similar effect
occurs. For a $45^\circ$-tilting of the order parameter no real-valued
admixture develops due to the symmetry of the system.

In conclusion we have shown that, for a $45^\circ$-rotated order parameter,
surface roughness suppresses the subdominant admixture in the
$d_{xy}$-wave as well as in the $s$-wave case. We have also found that the
subdominant component may vanish completely for very strong roughness.
As expected the $d_{xy}$-wave component is more strongly affected by
roughness than an $s$-wave component.  For a tilting angle
$\alpha<45^\circ$, a real-valued admixture occurs as well; the order
parameter is modified so that its suppression is weakened.  Our
results imply that surface roughness plays a crucial role for the
experimental observation of subdominant pairing. In particular the
roughness must be small enough to detect the splitting of the
zero-bias conductance peak.

\begin{acknowledgement}
  Helpful discussions with U. Eckern and P. Schwab are gratefully
  acknowledged. This work was supported by the Deutsche
  Forschungsgemeinschaft (project: LU 863/1-1).
\end{acknowledgement}


\begin{thebibliography}{10}

\bibitem{Ma95}
M.~Matsumoto and H.~Shiba,
\newblock J.\ Phys.\ Soc.\ Jpn. {\bf 64}, 3384 (1995).

\bibitem{Ma951}
M.~Matsumoto and H.~Shiba,
\newblock J.\ Phys.\ Soc.\ Jpn. {\bf 64}, 4867 (1995).

\bibitem{Fo97}
M.~Fogelstr{\"o}m, D.~Rainer, and J.~A. Sauls,
\newblock Phys.\ Rev.\ Lett. {\bf 79}, 281 (1997).

\bibitem{Fo03}
M.~Fogelstr{\"o}m, D.~Rainer, and J.~A. Sauls,
\newblock cond-mat/0302197.

\bibitem{Co97}
M.~Covington {\em et~al.},
\newblock Phys.\ Rev.\ Lett. {\bf 79}, 277 (1997).

\bibitem{Da01}
Y.~Dagan and G.~Deutscher,
\newblock Phys.\ Rev.\ Lett. {\bf 87}, 177004 (2001).

\bibitem{Al97}
L.~Alff {\em et~al.},
\newblock Phys.\ Rev.\ B {\bf 55}, 14757 (1997).

\bibitem{Ig00}
I.~Iguchi, W.~Wang, M.~Yamazaki, Y.~Tanaka, and S.~Kashiwaya,
\newblock Phys.\ Rev.\ B {\bf 62}, R6131 (2000).

\bibitem{Ye01}
N.-C. Yeh {\em et~al.},
\newblock Physica C {\bf 364-365}, 450 (2001).

\bibitem{Be98}
J.~J. Betouras and R.~Joynt,
\newblock Phys.\ Rev.\ B {\bf 57}, 11752 (1998).

\bibitem{Ba97}
D.~B. Bailey, M.~Sigrist, and R.~B. Laughlin,
\newblock Phys.\ Rev.\ B {\bf 55}, 15239 (1997).

\bibitem{Ei68}
G.~Eilenberger,
\newblock Z.\ Phys. {\bf 214}, 195 (1968).

\bibitem{RaSm86}
J.~Rammer and H.~Smith,
\newblock Rev.\ Mod.\ Phys. {\bf 58}, 323 (1986).

\bibitem{TL01}
T.~L{\"u}ck, U.~Eckern, and A.~Shelankov,
\newblock Phys.\ Rev.\ B {\bf 63}, 64510 (2001).

\bibitem{Za84}
A.~Zaitsev,
\newblock Sov.\ Phys.\ JETP {\bf 59}, 1015 (1984).

\bibitem{Ya96}
K.~Yamada, Y.~Nagato, S.~Higashitani, and K.~Nagai,
\newblock J. Phys. Soc. Jpn. {\bf 65}, 1540 (1996).

\bibitem{BaSv97}
Y.~S. Barash, A.~A. Svidzinski, and H.~Burkhardt,
\newblock Phys.\ Rev.\ B {\bf 55}, 15239 (1997).

\bibitem{Po99}
A.~Poenicke, Y.~S. Barash, C.~Bruder, and V.~Istyukov,
\newblock Phys.\ Rev.\ B {\bf 59}, 7102 (1999).

\bibitem{Po00}
A.~Poenicke, M.~Fogelstr{\"o}m, and J.~A. Sauls,
\newblock Physica\ B {\bf 284-288}, 589 (2000).

\bibitem{ShOz00}
A.~Shelankov and M.~Ozana,
\newblock Phys.\ Rev.\ B {\bf 61}, 7077 (2000).

\bibitem{TL03}
T.~L{\"u}ck, P.~Schwab, U.~Eckern, and A.~Shelankov,
\newblock cond-mat/0310250.

\bibitem{Ra98}
D.~Rainer, H.~Burkhardt, M.~Fogelstr{\"o}m, and J.~A. Sauls,
\newblock J.\ Phys.\ Chem. Solids {\bf 59}, 2040 (1998).

\end{thebibliography}
\end{document}